\documentclass[10pt,twocolumn,letterpaper]{article}

\usepackage[pagenumbers]{cvpr} 

\usepackage{graphicx}
\usepackage{amsmath}
\usepackage{amssymb}
\usepackage{booktabs}
\usepackage{float}
\usepackage{caption}
\usepackage{colortbl}
\usepackage{xcolor}
\usepackage{float}
\usepackage[export]{adjustbox}
\usepackage[symbol]{footmisc}

\usepackage[pagebackref,breaklinks,colorlinks]{hyperref}

\usepackage[capitalize]{cleveref}
\crefname{section}{Sec.}{Secs.}
\Crefname{section}{Section}{Sections}
\Crefname{table}{Table}{Tables}
\crefname{table}{Tab.}{Tabs.}

\newcommand{\bM}{\mathbf{M}}

\newcommand{\br}{\mathbf{r}}

\newcommand{\bT}{\mathbf{T}}

\newcommand{\bX}{\mathbf{X}}
\newcommand{\bY}{\mathbf{Y}}
\newcommand{\bZ}{\mathbf{Z}}

\newcommand{\nR}{\mathbb{R}}

\newcommand{\figref}[1]{Fig.~\ref{#1}}

\newcommand{\eqnref}[1]{Eq.~\eqref{#1}}
\newcommand{\tabnref}[1]{Table~\ref{#1}}

\makeatletter
\DeclareRobustCommand\onedot{\futurelet\@let@token\@onedot}
\def\@onedot{\ifx\@let@token.\else.\null\fi\xspace}
 
\def\ie{i.e\onedot}

\makeatother

\newcommand{\PAR}[1]{\vspace{0.1cm}\noindent{\bf #1} }

\def\cvprPaperID{11297} 
\def\confName{CVPR}
\def\confYear{2024}

\begin{document}
	\title{SCINeRF: Neural Radiance Fields from a Snapshot Compressive Image}
	
	\author{Yunhao Li$^{1,2}$\qquad Xiaodong Wang$^{1,2}$\qquad Ping Wang$^{1,2}$\qquad Xin Yuan$^{2}$\qquad Peidong Liu$^{2}$\footnotemark[2] \\$^{1}$Zhejiang University\qquad $^{2}$Westlake University\\
		{\tt\small \{liyunhao, wangxiaodong, wangping, xyuan, liupeidong\}@westlake.edu.cn}}
	
	\twocolumn[{
		\maketitle
		\vspace{-2.em}
		\begin{center}
			\setlength\tabcolsep{1pt}
			\begin{tabular}{c}
				\includegraphics[width=1.0\textwidth]{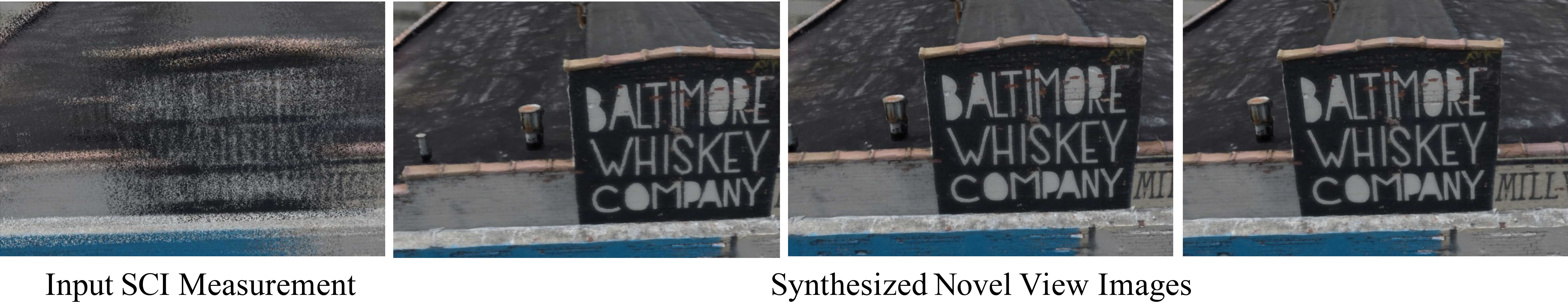}
			\end{tabular}
			\vspace{-1.em}
			\captionof{figure}{Given a single snapshot compressed image, our method is able to recover the underlying 3D scene representation. Leveraging the strong novel-view image synthesis capabilities of NeRF, we can render multi-view consistent images in high quality from {\bf the single measurement}.}
            \vspace{-0.5em}
			\label{fig_teaser}
		\end{center}
		}]
		
	\begin{abstract}
\vspace{-0.8em}
In this paper, we explore the potential of Snapshot Compressive Imaging (SCI) technique for recovering the underlying 3D scene representation from a single temporal compressed image. SCI is a cost-effective method that enables the recording of high-dimensional data, such as hyperspectral or temporal information, into a single image using low-cost 2D imaging sensors. To achieve this, a series of specially designed 2D masks are usually employed, which not only reduces storage requirements but also offers potential privacy protection. 
Inspired by this, to take one step further, our approach builds upon the powerful 3D scene representation capabilities of neural radiance fields (NeRF). Specifically, we formulate the physical imaging process of SCI as part of the training of NeRF, allowing us to exploit its impressive performance in capturing complex scene structures. To assess the effectiveness of our method, we conduct extensive evaluations using both synthetic data and real data captured by our SCI system. Extensive experimental results demonstrate that our proposed approach surpasses the state-of-the-art methods in terms of image reconstruction and novel view image synthesis. 
Moreover, our method also exhibits the ability to restore high frame-rate multi-view consistent images by leveraging SCI and the rendering capabilities of NeRF. The code is available at \texttt{\textcolor{magenta}{https://github.com/WU-CVGL/SCINeRF}}.
%
\vspace{-1.5em}
\end{abstract}    
	\section{Introduction}
\label{sec:intro}

Conventional high-speed imaging systems often face challenges such as high hardware cost and storage requirement. Drawing inspiration from Compressed Sensing (CS)\cite{Candes2006TIT, Donoho2006TIT}, video Snapshot Compressive Imaging (SCI)~\cite{yuan2021snapshot} system has emerged to address these limitations. A conventional video SCI system usually contains a hardware encoder and a software decoder. The hardware encoder employs a series of specially designed 2D masks to modulate the incoming photons across exposure time into a single compressed image. It enriches the low-cost cameras the ability to capture high-speed scenes, which further reduces the storage requirement. The whole encoding process can also be achieved via software implementation on pre-captured images, which can reduce the storage/transmission requirement and offer additional privacy protection. On the other hand, the software decoder restores the high frame-rate images using the compressed measurement and corresponding binary masks.

In recent years, several image reconstruction algorithms have been proposed  for SCI, which range from the model-based methods \cite{yuan2015generalized, liao2014generalized, liu2018rank} to deep-learning based methods \cite{qiao2020deep, cheng2020birnat, cheng2021memory, ma2019deep, wang2022spatial, wang2023efficientsci}. These algorithms can reconstruct encoded images or video frames with relatively high quality. However, these methods usually do not consider the underlying 3D scene structure to ensure multi-view consistency and can only recover images corresponding to the applied encoding masks. To address these challenges, we propose SCINeRF which recovers the underlying 3D scene representation from a single compressed image. High frame-rate multi-view consistency images can then be rendered from the learned 3D representation, as shown in Fig. \ref{fig_teaser}.

Our method exploits neural radiance fields (NeRF) \cite{mildenhall2021nerf} to represent the 3D scene implicitly. Different from prior explicit 3D representation techniques, NeRF takes pixel location and its corresponding camera pose as input, and predicts the pixel intensity by volume rendering. This characteristic makes it well suited for modeling the pixel-wise modulation process of an SCI system. Since it is impossible to recover the camera poses from a single compressed image via COLMAP \cite{schonberger2016structure}, we conduct a joint optimization on both the camera poses and NeRF, via minimizing the difference between the synthesized compressed image and real measurement from the encoder. Since SCINeRF performs test-time optimization, it does not suffer from the generalization performance degradation as prior end-to-end deep learning based methods. With the help of our proposed SCINeRF, we can capture the scenes within a single exposure time (can be shorter than 20ms or even 10ms) from a fast-moving camera, then recover the underlying 3D scene representations.

To better evaluate the performance of our method, we setup {\bf a real hardware platform} to collect real snapshot compressed images. We also conduct quantitative evaluations using synthetic images generated via Blender. Experimental results demonstrate that SCINeRF achieves superior performance over previous state-of-the-art methods in terms of image restoration and novel view image synthesis. In summary, our contributions are list as follows.
\vspace{-0.3em}
\begin{itemize}
	\itemsep0em
    \item We present a novel method to restore 3D aware multi-view images from a single snapshot compressed image, under the framework of NeRF. 
    \item We experimentally validate that our approach is able to synthesize high quality novel view images, surpassing existing state-of-the-art SCI image/video reconstruction methods.
    \item SCINeRF also presents an alternative approach for both efficient and privacy protection transmission between the edge devices and cloud infrastructure for practical NeRF deployment.  
\end{itemize}

	\section{Related Work}
\label{sec:related}
We review two main areas of the prior works: snapshot compressive imaging and NeRF, which are the most related components in our work.

\textbf{Snapshot Compressive Imaging.} Early SCI image decoding/reconstruction methods focus on regularized optimization-based approach \cite{yang2020shearlet, liao2014generalized, yuan2015generalized, liu2018rank}. It estimates the input compressed images from SCI encoding system by solving an optimization problem iteratively. In SCI, different regularizers and priors have been used, including sparsity \cite{yang2020shearlet} and total variation (TV) \cite{yuan2015generalized}. When solving the optimization problems, instead of using conventional gradient descent algorithm, most of the existing methods employ alternating direction method of multipliers (ADMM) \cite{boyd2011distributed}, which leads to good results and easier to adapt to different systems. The decompress SCI (DeSCI) \cite{liu2018rank} and GAP-TV \cite{yuan2015generalized} are state-of-the-art optimization-based approaches. However, this type of methods suffer from long running time and low flexibility to high resolution images. 

With the rise of deep learning and neural networks, most of the recent SCI decoding/reconstruction methods are based on deep learning. Various kinds of network architectures have been adopted into SCI decoder, including U-net \cite{ronneberger2015u}, and generative adversarial networks (GAN) \cite{NIPS2014_5ca3e9b1}. These deep learning-based methods takes thousands or even millions of synthetic SCI measurements and masks as training dataset (since obtaining large amount of real SCI measurement was time consuming and extremely difficult), and optimize the network using various kinds of loss functions including mean square error (MSE), feature loss \cite{johnson2016perceptual} and GAN loss \cite{miao2019net}. Qiao et al. \cite{qiao2020deep} built an End-to-End CNN (E2E-CNN) with reconstruction loss to retrieve compressed images. Cheng et al. \cite{cheng2020birnat} employed bidirectional recurrent neural networks (BIRNAT), which includes a bi-directional residual neural network (RNN) \cite{he2016deep} to reconstruct the temporal frames in a sequential manner. RevSCI \cite{cheng2021memory} reduced the time and memory complexity during large-scale video SCI training by designing a multi-group reversible 3D CNN architecture. ADMM-Net \cite{ma2019deep} modeled the decoding process as a tensor recovery problem from random linear measurements and interprets the ADMM process into a deep neural network. MetaSCI \cite{wang2021metasci} adopted a meta modulated CNN to make the reconstruction network more adaptable to large scale data and novel masks. Plug-and-play fast and flexible denoising CNN (PnP-FFDNet) \cite{yuan2020plug} combined deep denoising network with ADMM, leading to the fast and flexible reconstruction. Later, Yuan et al. \cite{yuan2021plug} developed fast deep video denoising network (FastDVDNet), which improves the performance of PnP-FFDNet by introducing the the most recent deep denoising network.  Wang et al. introduced spatial-temporal Transformer (STFormer) \cite{wang2022spatial} and EfficientSCI \cite{wang2023efficientsci} to exploit the spatial and temporal correlation within image decoding process utilizing Transformer \cite{vaswani2017attention} architecture. These deep learning-based methods can reconstruct  images with relatively high quality. However, since these methods contain pre-trained models from synthetic datasets, they lack generalization abilities and perform poor on real datasets. Moreover, existing deep learning-based methods can only reconstruct images corresponding to the masks, making it difficult to restore high frame-rate images and videos. On the contrary, our method performs test-time optimization with no generalization problems, and it can restore high frame-rate multi-view images from estimated 3D scenes.   

\textbf{NeRF and Its Variants.} Mildenhall et al. \cite{mildenhall2021nerf, tancik2020fourier} proposed NeRF, an epochal novel scene representation method. NeRF estimates the scene appearance and geometry continuously and its performance surpasses state-of-the-art scene representation schemes by producing better novel-view images. To make NeRF robust to more complicated real-world environments, researchers have proposed many dedicated NeRF variants recently. Some researches extend NeRF to make it compatible for large scale scene reconstruction and representation \cite{gu2021stylenerf2, xiangli2022bungeenerf, turki2022mega, tancik2022block}. Others focus on developing NeRF which can deal with non-rigid object reconstruction\cite{athar2022rignerf, gafni2021dynamic, peng2021animatable, peng2021neural}. There are also many NeRF-based frameworks for high dynamic range (HDR) image modeling \cite{Huang_2022_CVPR}, scene editing \cite{yang2021learning, yuan2022nerf, weder2023removing, wei2023clutter} and scene appearance decomposition\cite{bi2020neural, boss2021nerd}.

We will mainly review NeRF-based methods close to our work in detail in this section. The original NeRF requires the accurate camera poses as input. Since many existing image datasets lack camera pose information, many NeRF-based methods, including the original NeRF framework, estimate the camera poses from input images. They employ widely used structure-from-motion (SfM) software COLMAP \cite{schonberger2016structure} to get camera poses. But in many cases, the camera poses estimated from SfM are unavailable or inaccurate. To address this issue, many frameworks focus on NeRF optimization without camera poses. Wang et al. propose NeRF-- \cite{wang2021metasci}, which jointly estimates the scene and camera parameters at each training iteration. Jeong \cite{jeong2021self} develops a NeRF-based camera self-calibration scheme. iMAP \cite{sucar2021imap}, introduced by Sucar et al., integrates NeRF into Simultaneous Localization and Mapping (SLAM) to optimize the scene and camera trajectory together. Meng et al. \cite{meng2021gnerf} implement GAN to solve the problem of optimizing NeRF without camera poses. Other researches focus on refining inaccurate camera poses during the NeRF optimization process. Bundle adjusting neural radiance fields (BARF) \cite{lin2021barf} proposed by Lin et al. employs a simple coarse-to-fine registration strategy, which applies a smooth mask on the positional encoding serving as a dynamic low-pass filter. This strategy allows BARF to optimize the scene representation with camera poses. Following the idea of BARF, Wang et al. \cite{Wang_2023_CVPR} propose bundle-adjusted deblur NeRF (BAD-NeRF), which optimizes the camera trajectories and NeRF network together by linearly interpolating camera poses to imitate the image formation model of motion blur, so that it can estimate the clear scenes from blurred images.

	\section{Method}
\label{sec:method}

\begin{figure*}[!htbp]
  \centering
    \includegraphics[width=0.99\linewidth]{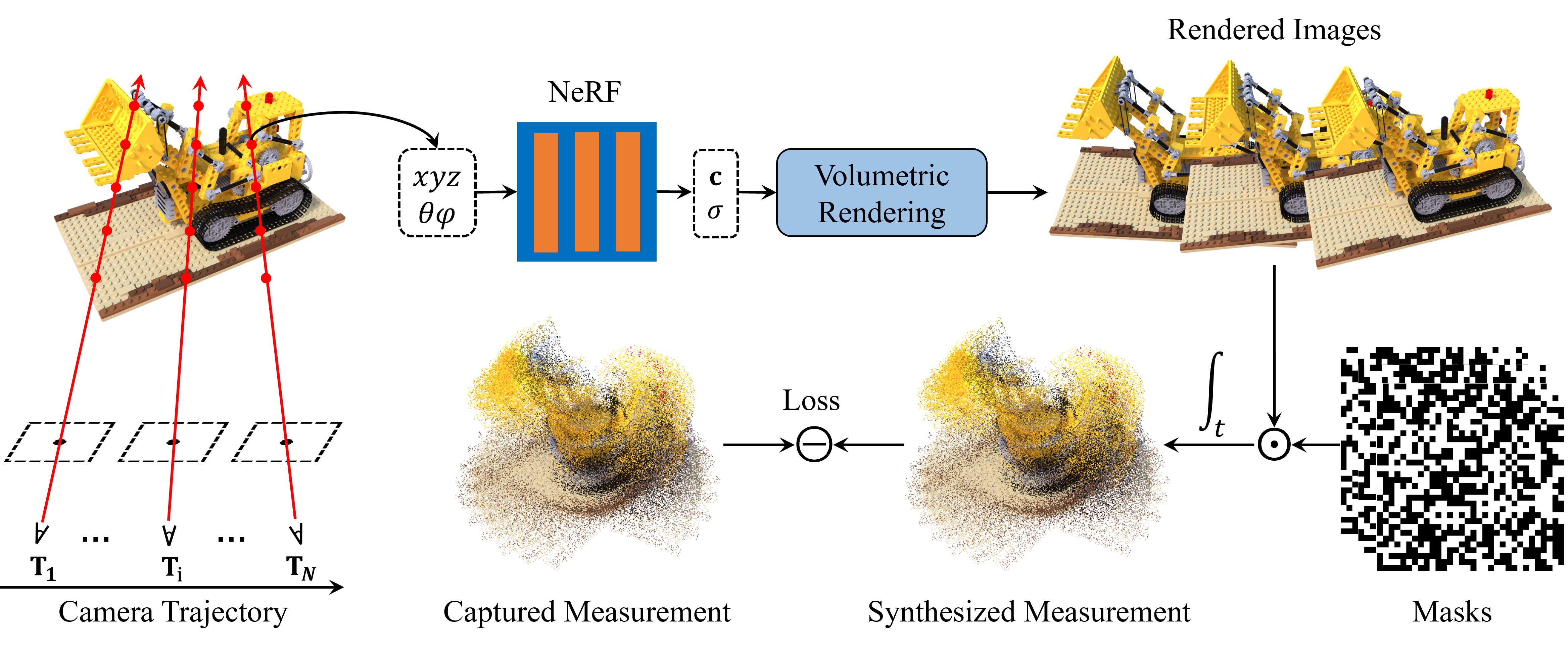}
    \caption{\textbf{Overview of the proposed SCINeRF.} Our method takes a single snapshot compressed image and corresponding masks as input, and recovers the underlying 3D scene representation as well as the camera motion trajectory within a single exposure time.}
    \label{fig2}
\end{figure*}

Our method takes a single compressed image and encoding masks as input, and recovers the underlying 3D scene representation as well as camera poses. High frame-rate images can then be rendered from the learned 3D scene representation. To achieve this, we exploit NeRF \cite{mildenhall2021nerf} as the underlying 3D scene representation due to its impressive representation capability.  We follow the real image formation process to synthesize snapshot compressed image from NeRF. By maximizing the photometric consistency between the synthesized image and the actual measurements, we optimize both NeRF and the camera poses. \figref{fig2} presents the overview of our method. We detail each component as follows.

\subsection{Background on NeRF} 
Given a set of input multi-view images (together with both the camera intrinsic and extrinsic parameters), NeRF \cite{mildenhall2021nerf} transfers the pixels of the input images into rays. It then samples points along each ray, and takes 5D vectors (i.e., the 3D position of sampled point and the 2D viewing directions) as input. The volume density $\sigma$ and view-dependent RGB color $\textbf{c}$ of each sampled 3D point are then estimated by a Multi-layer Perceptron (MLP). The reason that NeRF predicts color from both position and viewing direction is to better deal with the specular reflection of the scene. After obtaining the volume density and color of each sampled point along the ray, it employs a conventional volumetric rendering technique to integrate the density and color to synthesize the corresponding pixel intensity $C$ of the image. The whole process can be formally defined via following equation:
\begin{equation}
  C(\textbf{r})=\int_{t_n}^{t_f}T(t)\sigma(\textbf{r}(t))\textbf{c}(\textbf{r}(t),\textbf{d})dt,
\end{equation}
where $t_n$ and $t_f$ are near and far bounds for volumetric rendering respectively, $\textbf{r}(t)$ is the sampled 3D point along the ray $\textbf{r}$ at the distance $t$ from the camera center,  $\sigma(\textbf{r}(t))$ represents the predicted density of the sampled point $\textbf{r}(t)$ by the MLP, $T(t)$ denotes the accumulated transmittance along the ray from $t_n$ to $t$, and is defined as $\exp(-\int_{t_n}^{t}T(t)\sigma(\textbf{r}(s))ds)$, $\textbf{d}$ is the viewing direction in the world coordinate frame, and $\textbf{c}(\textbf{r}(t),\textbf{d})$ is the predicted color of the sampled point $\textbf{r}(t)$ by the MLP.

The photometric loss, \ie the mean squared error (MSE) between the rendered pixel intensity and the real captured intensity, is usually used to train the networks:
\begin{equation}
    \mathcal{L}=\sum_{\textbf{r}\in\mathcal{R}}\left\|\hat{C}(\textbf{r})-C(\textbf{r})\right\|^{2},
\end{equation}
where both $C(\textbf{r})$ and $\hat{C}(\textbf{r})$ denote the rendered and real captured pixel intensities for ray \textbf{r} respectively, and $\mathcal{R}$ denotes the set of sampled rays.

\subsection{Image Formation Model of Video SCI}
The formation process of a video SCI system is similar to that of a blurry image. The difference is that the captured image of a video SCI system is modulated by $N$ binary masks $\{\mathbf{M}_i\}_{i=1}^{N}\in\mathbb{R}^{H\times W}$ across the exposure time, where both $H$ and $W$ are image height and width respectively. For practical hardware implementation, those masks are achieved by displaying different 2D patterns on the Digital Micro-mirror Device (DMD) and a spatial light modulator, e.g., liquid crystal. The image sensor then accumulates the modulated photons across exposure time to a compressed/coded image. The number of masks or different patterns on the DMD within exposure time determine the number of coded frames, i.e., the temporal compression ratio (CR). Due to the mask modulation, the image restoration problem is not ill-posed anymore, \ie, the $N$ virtual images can be recovered from a single compressed image alone. 
The whole imaging process can be described formally as:
\begin{equation}\label{eq_formation}
	\mathbf{Y}=\sum_{i=1}^{N}\mathbf{X_i}\odot\mathbf{M_i}+\mathbf{Z},
\end{equation}
where $\bY, \bX_i \in \nR^{H \times W}$ are the captured compressed image and the $i^{th}$ virtual image within exposure time respectively, $N$ is the temporal CR, $\odot$ denotes element-wise multiplication, and $\bZ \in \nR^{H \times W}$ is the measurement noise. The individual pixel value in the binary mask is randomly generated. For $N$ masks across the exposure time, the probability of assigning 1 to the same pixel location is fixed and defined as overlapping ratio. The optimal overlapping ratio is selected via ablation study. 

Given the NeRF representation and corresponding camera poses, we are able to render $\bX_i$ to synthesize the compressed image $\bY$. We can see that $\bY$ is differentiable with respect to NeRF and the poses, which lay the foundation for our optimization step.

\subsection{Proposed Framework} 
Based on \eqnref{eq_formation}, we can model the SCI measurement (\ie $\bY$) as the compressed multi-view images. However, we cannot estimate the scene from SCI measurement directly because of lack of camera poses. Since the SCI measurement contains compressed frames and it is in the form of 2D image, it is unrealistic to use the conventional SfM method to estimate the camera pose for each frame, as most of the NeRF-based methods. 

To deal with the unknown poses problem, we follow the idea of prior methods \cite{lin2021barf,wang2021nerf,Wang_2023_CVPR}, which start from the initialized inaccurate poses and optimize the parameters of NeRF together with poses simultaneously. Since the compressed multi-view images are taken within a relatively-short exposure time, we assume that the camera trajectory during the imaging process is linear and obtain poses using linear interpolation. For more complex motions, we can exploit higher-order spline \cite{li2023usbnerf} or directly optimize individual poses without the loss of generality. The virtual camera pose can thus be represented as:
\begin{equation}
    \mathbf{T_i} = \mathbf{T_1}\exp({\frac{i}{N}\log(\mathbf{T_1^{-1}T_N})}),
\end{equation}
where $\mathbf{T_i}$ is the pose of $i^{th}$ frame, $N$ denotes the number of compressed frames, $\mathbf{T_1}\in \mathbf{SE}(3)$ and $\mathbf{T_N}\in \mathbf{SE}(3)$ are the poses of the first and last frame respectively. They represent start and end pose of the camera trajectory. Both $\bT_1$ and $\bT_N$ are initialized to be nearly identity matrix with small random perturbations to the translation components. The intrinsic parameters, including the camera focal length and principle points, come from camera intrinsic calibration, which is a well-studied problem with abundant methods available. 

Given the image formation model and the camera pose representations, we are now able to jointly optimize both the NeRF parameters and camera poses by minimizing following loss function:
\begin{equation}
    \mathcal{L}=\sum_{\textbf{r}\in\mathcal{R}}\left\|\hat{\bY}(\textbf{r})-\sum_{i=1}^{N}\mathbf{M}(\br,i) \odot C(\textbf{r},i)\right\|^{2}, 
\end{equation}
where $\mathcal{R}$ denotes the set of sampled rays $\textbf{r}$, $\hat{\bY}(\textbf{r})$ is pixel value of the real captured image corresponding to $\br$, $\bM(\br,i)$ is the binary mask value of the $i^{th}$ mask corresponding to ray $\br$, $\odot$ represents element-wise multiplication, and $C(\textbf{r},i)$ denotes the rendered pixel value of the $i^{th}$ frame corresponding to $\br$.

	\section{Experiments}

\begin{table*}
	\setlength\tabcolsep{2pt}
	\parbox{\textwidth}{
		\resizebox{\linewidth}{!}{
		\begin{tabular}{c|ccc|ccc|ccc|ccc|ccc|ccc}
			\specialrule{0.1em}{1pt}{1pt}
			& \multicolumn{3}{c|}{Airplants} & \multicolumn{3}{c|}{Hotdog} & \multicolumn{3}{c|}{Cozy2room} & \multicolumn{3}{c|}{Tanabata} & \multicolumn{3}{c|}{Factory} & \multicolumn{3}{c}{Vender} \\
			& PSNR$\uparrow$ & SSIM$\uparrow$ & LPIPS$\downarrow$ & PSNR$\uparrow$ & SSIM$\uparrow$ & LPIPS$\downarrow$ & PSNR$\uparrow$ & SSIM$\uparrow$ & LPIPS$\downarrow$ & PSNR$\uparrow$ & SSIM$\uparrow$ & LPIPS$\downarrow$ & PSNR$\uparrow$ & SSIM$\uparrow$ & LPIPS$\downarrow$ & PSNR$\uparrow$ & SSIM$\uparrow$ & LPIPS$\downarrow$ \\
			\specialrule{0.05em}{1pt}{1pt}
			GAP-TV \cite{yuan2015generalized} &22.85 &.4057 &.4986 & 22.35 &.7663 &.3179 & 21.77 &.4321 &.6031 & 20.42 &.4264 &.6250 & 24.05 &.5666 &.5149 &20.00 &.3678 &.6882\\
			PnP-FFDNet \cite{yuan2020plug} &27.79 &.9117 &.1817 &29.00 &.9765 &.0511 &28.98 &.8916 &.0984 & 29.17 & .9032 & .1197 & 31.75 & .8977 & .1142 & 28.70 & .9235 & .1315 \\
			PnP-FastDVDNet \cite{yuan2021plug} &28.18 & .9092 & .1757 & 29.93 & .9728 & .0522 & 30.19 & .9132 & .0793 & 29.73 & .9333 & .0980 & 32.53 & .9165 & .1055 & 29.68 & .9395 & .1043 \\
			EfficientSCI \cite{wang2023efficientsci} &30.13 & {\textbf{.9425}} & .1129 & 30.75 & .9568 & .0461 & 31.47 & .9327 & .0476 & 32.30 & .9587 & .0600 & 32.87 & .9259 & .0709 & 33.17 & .9401 & .0456 \\
			\specialrule{0.05em}{1pt}{1pt}
			ours &{\textbf{30.69}} &.9335 &{\textbf{.0728}} &{\textbf{31.35}} &{\textbf{.9878}} &{\textbf{.0310}} &{\textbf{33.23}} & {\textbf{.9492}} &{\textbf{.0445}} &{\textbf{33.61}} &{\textbf{.9638}} &{\textbf{.0374}} &{\textbf{36.60}} &{\textbf{.9638}} &{\textbf{.0221}} &{\textbf{36.40}} &{\textbf{.9840}} &{\textbf{.0298}}\\
			\specialrule{0.1em}{1pt}{1pt}
	\end{tabular}}
    \vspace{-0.7em}
	\caption{{\textbf{Quantitative SCI image reconstruction comparisons on the synthetic datasets}} The results are computed from the rendered images from estimated scenes via SCINeRF and recovered images from state-of-the-art SCI image restoration methods. The experimental results demonstrate that our SCINeRF can render images with higher quality than those from existing methods.}
	\label{table_2dt3d}}
\end{table*}

\begin{table*}
    \parbox{\linewidth}{
    \resizebox{\linewidth}{!}{
	\begin{tabular}{c|ccc|ccc|ccc|ccc|ccc|ccc}
		\specialrule{0.1em}{1pt}{1pt}
		& \multicolumn{3}{c|}{Airplants} & \multicolumn{3}{c|}{Hotdog} & \multicolumn{3}{c|}{Cozy2room} & \multicolumn{3}{c|}{Tanabata} & \multicolumn{3}{c|}{Factory} & \multicolumn{3}{c}{Vender} \\
		& PSNR$\uparrow$ & SSIM$\uparrow$ & LPIPS$\downarrow$ & PSNR$\uparrow$ & SSIM$\uparrow$ & LPIPS$\downarrow$ & PSNR$\uparrow$ & SSIM$\uparrow$ & LPIPS$\downarrow$ & PSNR$\uparrow$ & SSIM$\uparrow$ & LPIPS$\downarrow$ & PSNR$\uparrow$ & SSIM$\uparrow$ & LPIPS$\downarrow$ & PSNR$\uparrow$ & SSIM$\uparrow$ & LPIPS$\downarrow$ \\
		\specialrule{0.05em}{1pt}{1pt}
			NeRF+GAP-TV  &23.72 &.4684 &.4195 &23.80 &.7672 &.2847 & 21.99 &.5027 &.5212 & 20.91 &.4122 &.5531 &25.48 &.6853 &.4401 &21.68 &.4365 &.5536\\
			NeRF+PnP-FFDNet  &26.72 &.8831 &.2249 &29.14 &.9721 &.0618 &30.15 &.9030 &.0919 & 28.00 & .9025 & .1400 & 31.96 & .8850 & .1823 & 30.02 & .9343 & .1317 \\
			NeRF+PnP-FastDVDNet  &26.91 & .8766 & .2061 & 29.31 & .9729 & .0548 & 31.17 & .9170 & .0797 & 30.79 & .9362 & .1000 & 32.56 & .9026 & .1532 & 31.30 & .9451 & .1151 \\
			NeRF+EfficientSCI  &28.62 & .9140 & .1595 & 29.82 & .9766 & .0527 & 31.79 & .9357 & .0641 & 31.35 & .9535 & .0769 & 32.72 & .8942 & .0676 & 32.77 & .9431 & .0648 \\
		\specialrule{0.05em}{1pt}{1pt}
			ours &{\textbf{30.61}} &{\textbf{.9384}} &{\textbf{.0764}} &{\textbf{30.59}} &{\textbf{.9814}} &{\textbf{.0442}} &{\textbf{33.12}} & {\textbf{.9477}} &{\textbf{.0357}} &{\textbf{32.84}} &{\textbf{.9574}} &{\textbf{.0297}} &{\textbf{36.33}} &{\textbf{.9602}} &{\textbf{.0426}} &{\textbf{34.75}} &{\textbf{.9744}} &{\textbf{.0257}}\\
		\specialrule{0.1em}{1pt}{1pt}
	\end{tabular}}
    \vspace{-0.9em}
    \caption{{\textbf{Quantitative comparisons on novel-view synthesis}} The results are the rendered novel-view images from SCINeRF and novel-view images from vanilla NeRF with reconstructed images from existing state-of-the-art methods. Since we cannot estimate accurate poses from the SOTA outputs, we use ground truth images to estimate camera poses instead. The experimental results show that our SCINeRF outperforms existing approaches.}
    \label{table_3dt3d}}
	\vspace{-0.7em}
\end{table*}

\label{sec:exp}
We validate our SCINeRF on both synthetic and real datasets captured by our system, and it is evaluated against state-of-the-art (SOTA) SCI image restoration methods. The experimental results demonstrate that SCINeRF delivers higher performance compared to existing works in terms of restoration quality.

\subsection{Experimental Setup}
\PAR{Synthetic datasets.} We generate synthetic datasets from commonly-used multi-view datasets for image-based rendering and novel-view synthesis, including \textit{Airplants} in LLFF \cite{mildenhall2019local}  and \textit{Hotdog} in NeRF Synthetic360 \cite{mildenhall2021nerf}. We generate additional synthetic datasets from the scenes provided in DeblurNeRF \cite{ma2022deblur} with Blender, to better simulate the capturing process along a trajectory of real camera. There are four virtual scenes in the datasets synthesized by Blender, including \textit{Cozy2room, Tanabata, Factory,} and \textit{Vender}. For these four scenes, we generate images from 5 trajectories and compute the average image quality metrics. The compression ratio of benchmark datasets is 8. To verify the adaptability to different image resolutions of our SCINeRF, we use different spatial resolutions in synthetic datasets: For LLFF, we use 512 $\times$ 512, for NeRF Synthetic360 we use 400 $\times$ 400. For Blender-synthesized datasets, we use 600 $\times$ 400 resolution.

\begin{figure}[tbp]
  \centering
    \includegraphics[width=0.99\linewidth]{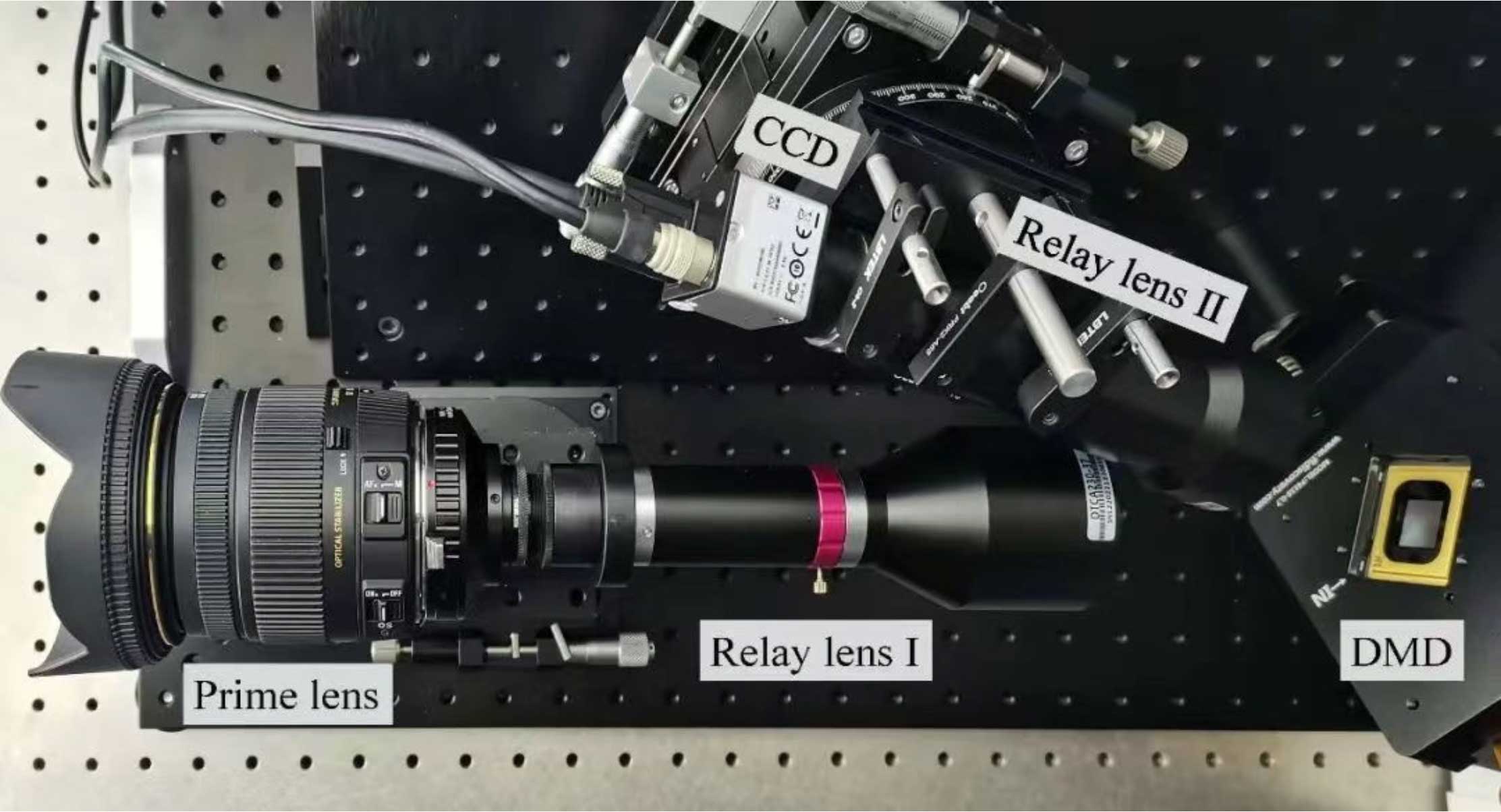}
    \caption{\textbf{Experimental setup for real dataset collection.} This SCI imaging system contains a CCD camera to record snapshot measurement, primary and rely lens, and a DMD to modulate input frames.}
    \label{experimental setup}
    \vspace{-0.8em}
\end{figure}

\PAR{Real-world datasets.} For real dataset, we collect the SCI measurement from a real video SCI setup. The setup consists of an iRAYPLE A5402MU90 camera and a FLDISCOVERY F4110 DMD. The compression ratio of the real dataset is 8, with the resolution of SCI measurement being 1024$\times$768. Fig. \ref{experimental setup} shows the experimental setup we used to collect real dataset.

\PAR{Baseline methods and evaluation metrics.} Since our SCINeRF can render high-quality images from the estimated scene, we compare our method against SOTA SCI image restoration methods, including GAP-TV \cite{yuan2015generalized}, PnP-FFDNet \cite{yuan2020plug}, PnP-FastDVDNet \cite{yuan2021plug}, and EfficientSCI \cite{wang2023efficientsci}. For fair comparisons, we fine-tuned EfficientSCI \cite{wang2023efficientsci} with the masks used in our datasets.
Additionally, we also compare the performance on novel view image synthesis of SCINeRF against that of vanilla NeRF with restored images from prior SOTA methods as input. We noticed that due to the lack of high-quality details and multi-view consistency, we cannot estimate camera poses via SfM using the images reconstructed from those SCI image reconstruction methods. So when feeding the reconstructed images into NeRF, we use camera poses estimated from ground truth images instead, making the comparison unfair to our SCINeRF (but providing reasonable results). 
Widely used metrics are employed for quantitative evaluations, such as the structural similarity index (SSIM), peak signal-to-noise-ratio (PSNR), and learned perceptual image patch similarity (LPIPS) \cite{zhang2018unreasonable}.

\PAR{Implementation details.} We implement our method with PyTorch \cite{lin2020nerfpytorch}. We exploit the vanilla NeRF from \cite{mildenhall2021nerf} to represent the 3D scene. For higher performance, more recent representation can be exploited \cite{mueller2022instant}. We use two separate Adam optimizers \cite{kingma2017adam} with learning rate for scene model decreasing from $5 \times 10^{-4}$ to $5\times10^{-5}$ exponentially, and the learning rate for pose optimization decreasing from $1\times10^{-3}$ to $1\times10^{-5}$ exponentially. We train our model for 100-200K iterations, with 5000 rays as batch size. 

\begin{figure*}[t]

\begin{minipage}[c]{1.0\textwidth}
\begin{minipage}[c]{\linewidth}
\centering
  \begin{minipage}[c]{0.155\linewidth}
  \centering
  \small
  \ \\  Measurement~
  \end{minipage}
  ~
  \begin{minipage}[c]{0.155\linewidth}
  \centering
  \small
  \ \\ GAP-TV~\cite{yuan2015generalized}
  \end{minipage}
  ~
  \begin{minipage}[c]{0.155\linewidth}
  \centering
  \small
  \ \\ \ PnP-FFDNet~\cite{yuan2020plug}
  \end{minipage}
  ~
  \begin{minipage}[c]{0.155\linewidth}
  \centering
  \small
  \ \\ \ EfficientSCI~\cite{wang2023efficientsci}
  \end{minipage}
  ~
  \begin{minipage}[c]{0.155\linewidth}
  \centering
  \small
  \ \\ \ Ours
  \end{minipage}
  ~
  \begin{minipage}[c]{0.155\linewidth}
  \centering
  \small
  \ \\ \ Ground Truth
  \end{minipage}
\end{minipage}
\\
\begin{minipage}[c]{\linewidth}
\centering
  \begin{minipage}[c]{0.155\linewidth}
  \includegraphics[width=1\linewidth]{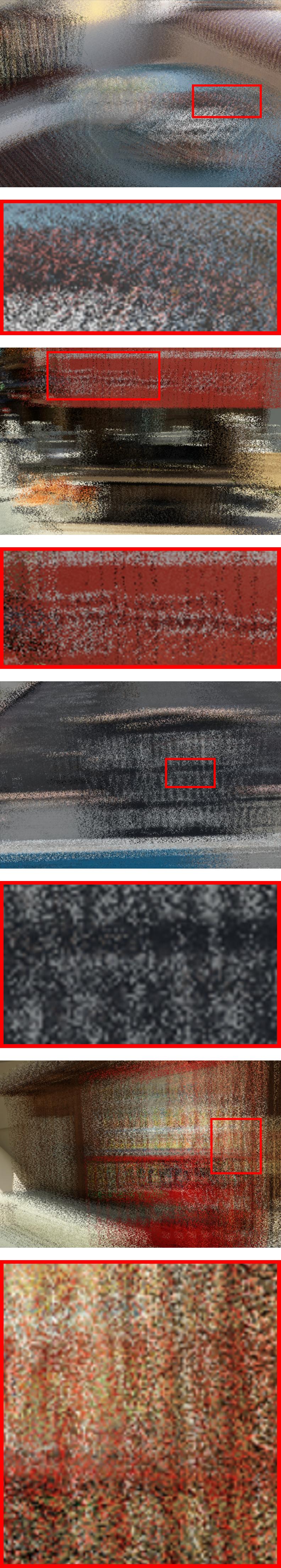}
  
  \end{minipage}
  ~
  \begin{minipage}[c]{0.155\linewidth}
  \includegraphics[width=1\linewidth]{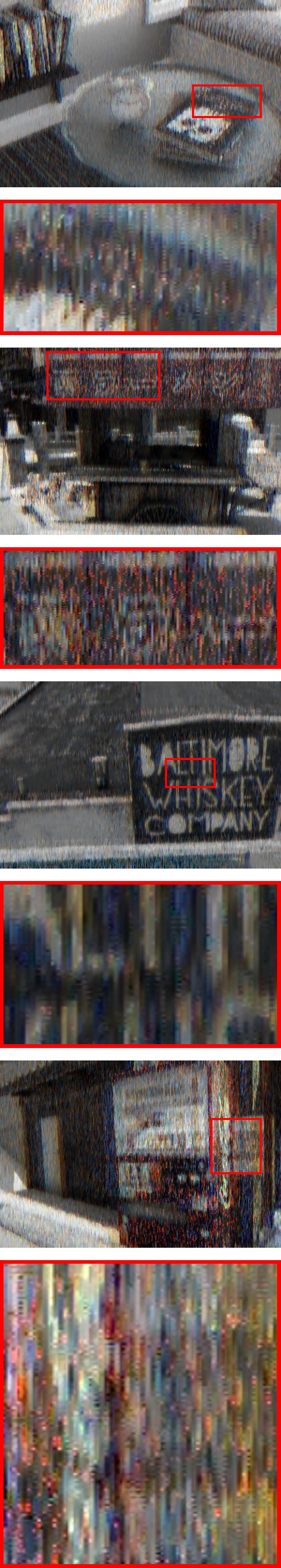}
  \end{minipage}
  ~
  \begin{minipage}[c]{0.155\linewidth}
  \includegraphics[width=1\linewidth]{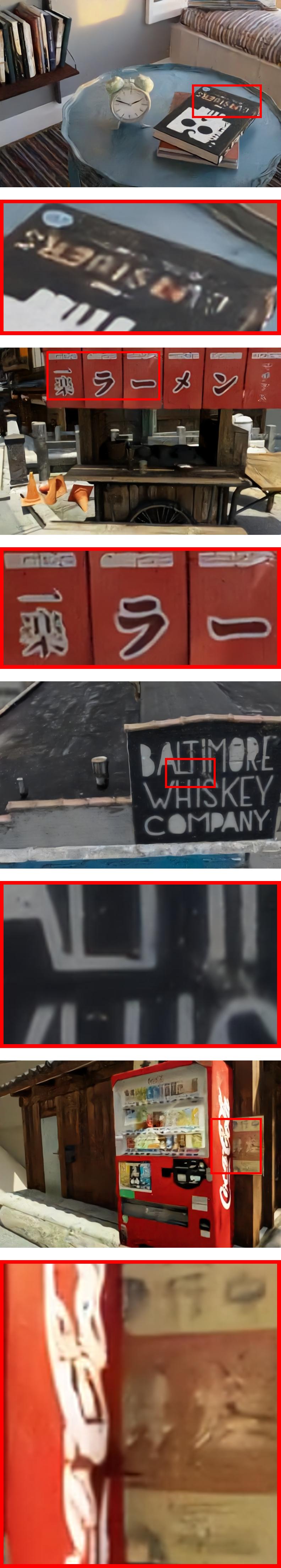}
  \end{minipage}
  ~
  \begin{minipage}[c]{0.155\linewidth}
  \includegraphics[width=1\linewidth]{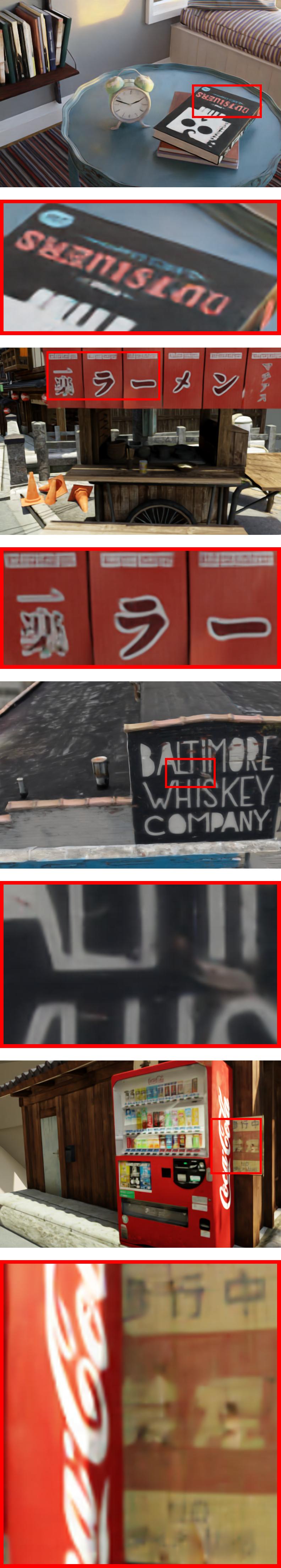}
  \end{minipage}
  ~
  \begin{minipage}[c]{0.155\linewidth}
  \includegraphics[width=1\linewidth]{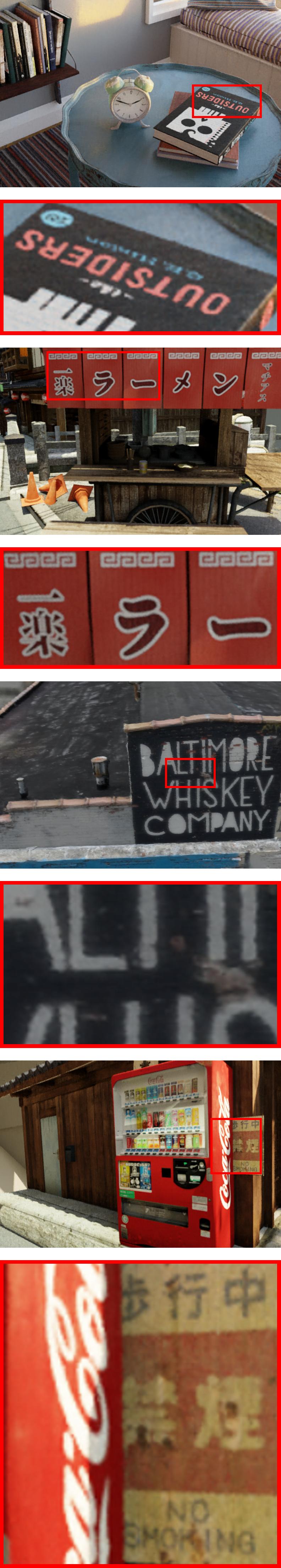}
  \end{minipage}
  ~
  \begin{minipage}[c]{0.155\linewidth}
  \includegraphics[width=1\linewidth]{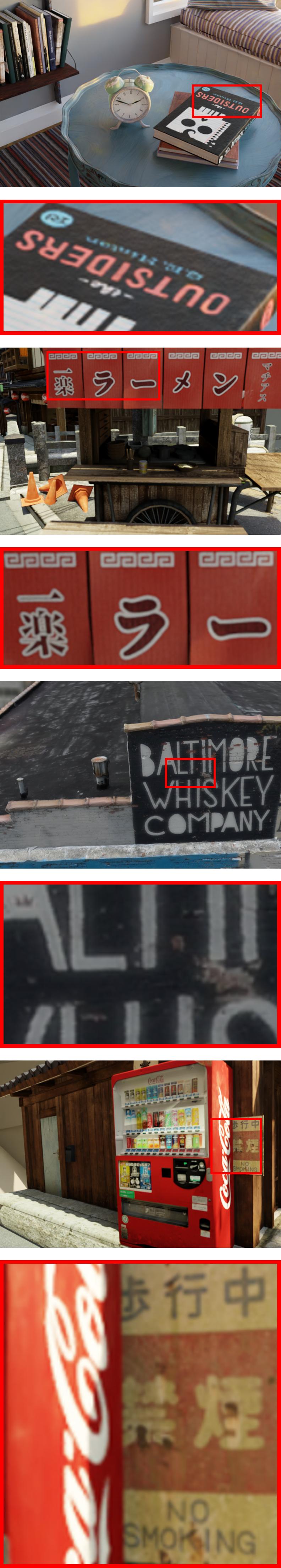}
  \end{minipage}
\end{minipage}
\end{minipage}
\\

   \caption{\textbf{Qualitative evaluations of our method against SOTA SCI image restoration methods on the synthetic dataset.} Top to bottom shows the results for different scenes, including \textit{Cozy2room}, \textit{Tanabata}, \textit{Factory} and \textit{Vender}. The experimental results demonstrate that our method achieves superior performance on image restoration from a single compressed image (the far-left column). 
   }
   \label{fig4}
    \vspace{-0.8em}
   
\end{figure*}
\begin{figure*}[t]

\begin{minipage}[c]{1.0\textwidth}
\begin{minipage}[c]{\linewidth}
\centering
  \begin{minipage}[c]{0.155\linewidth}
  \centering
  \small
  \ \\  Measurement~
  \end{minipage}
  ~
  \begin{minipage}[c]{0.155\linewidth}
  \centering
  \small
  \ \\ GAP-TV~\cite{yuan2015generalized}
  \end{minipage}
  ~
  \begin{minipage}[c]{0.155\linewidth}
  \centering
  \small
  \ \\ \ PnP-FFDNet~\cite{yuan2020plug}
  \end{minipage}
  ~
  \begin{minipage}[c]{0.155\linewidth}
  \centering
  \small
  \ \\ \ EfficientSCI~\cite{wang2023efficientsci}
  \end{minipage}
  ~
  \begin{minipage}[c]{0.155\linewidth}
  \centering
  \small
  \ \\ \ Ours
  \end{minipage}
  ~
  \begin{minipage}[c]{0.155\linewidth}
  \centering
  \small
  \ \\ \ Original Scene
  \end{minipage}
\end{minipage}
\\
\begin{minipage}[c]{\linewidth}
\centering
  \begin{minipage}[c]{0.155\linewidth}
  \includegraphics[width=1\linewidth]{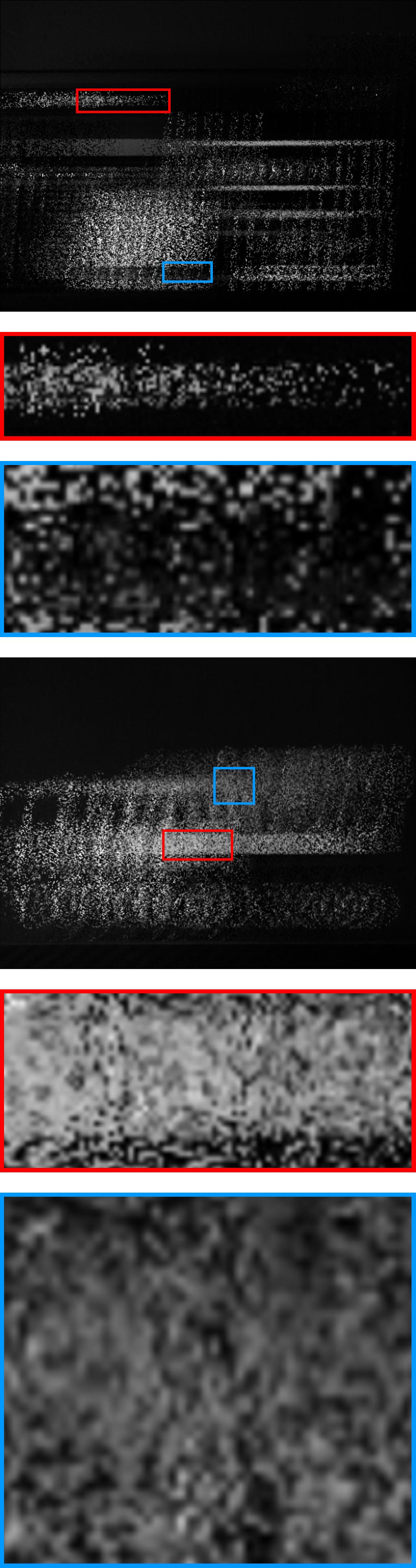}
  
  \end{minipage}
  ~
  \begin{minipage}[c]{0.155\linewidth}
  \includegraphics[width=1\linewidth]{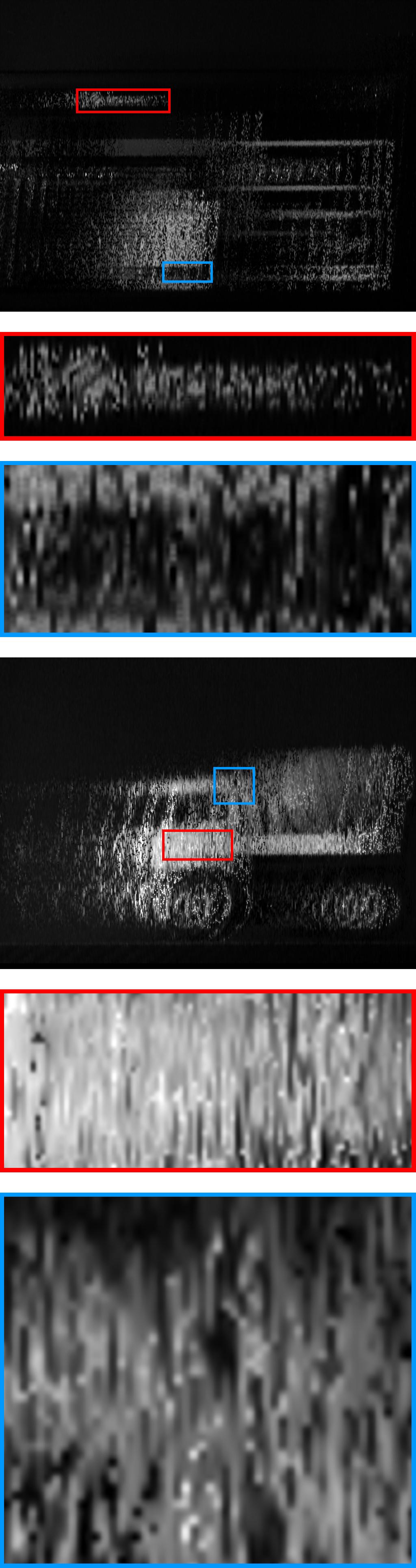}
  \end{minipage}
  ~
  \begin{minipage}[c]{0.155\linewidth}
  \includegraphics[width=1\linewidth]{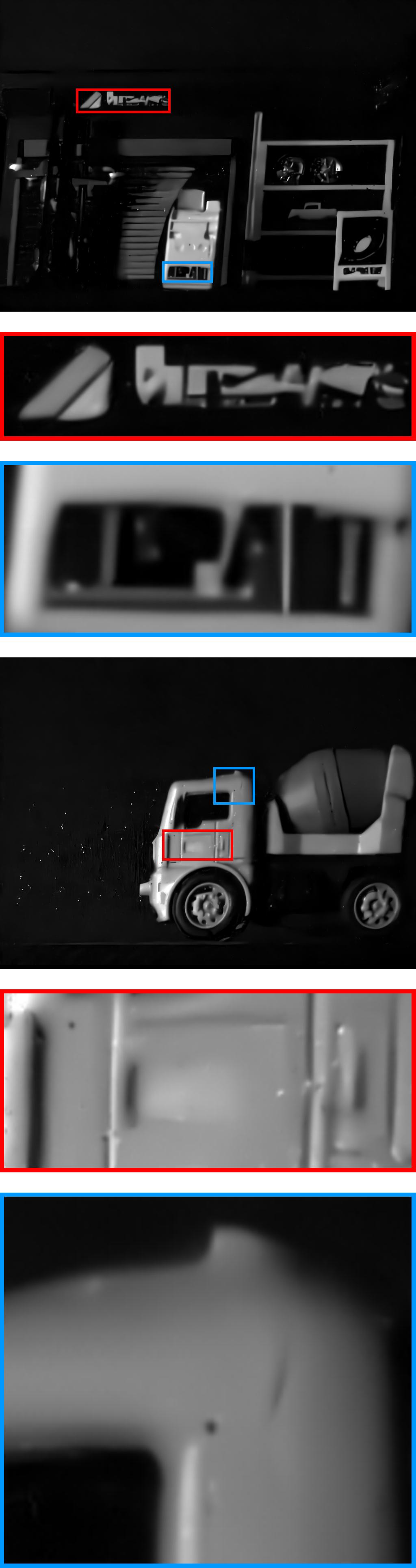}
  \end{minipage}
  ~
  \begin{minipage}[c]{0.155\linewidth}
  \includegraphics[width=1\linewidth]{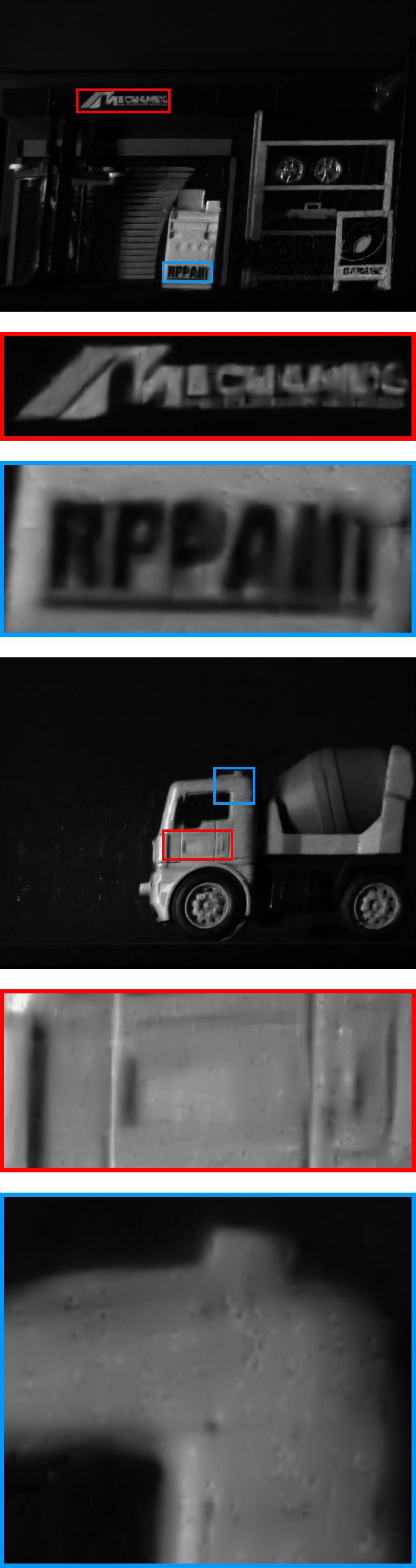}
  \end{minipage}
  ~
  \begin{minipage}[c]{0.155\linewidth}
  \includegraphics[width=1\linewidth]{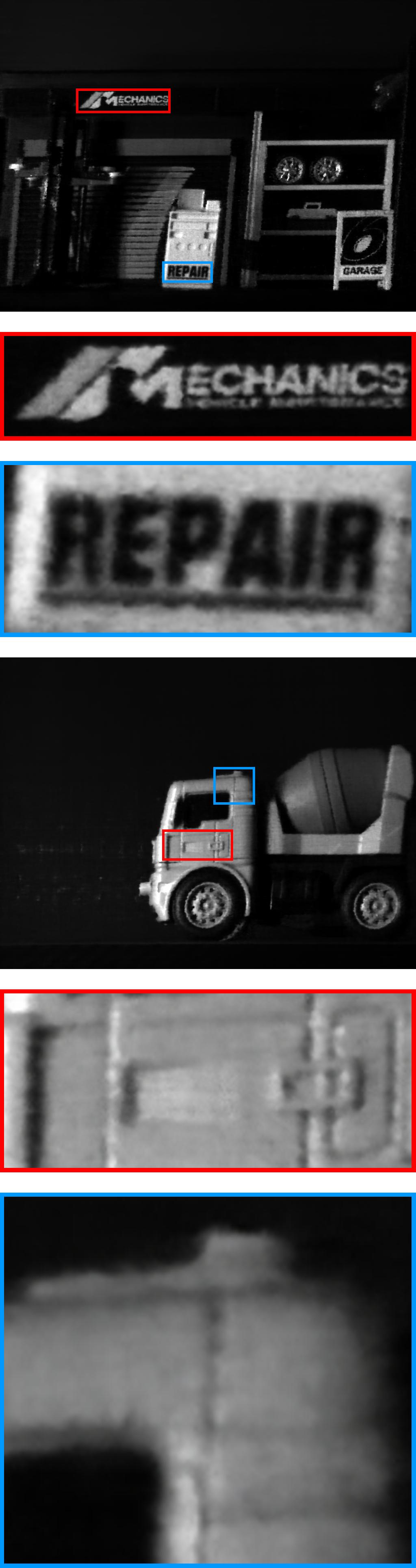}
  \end{minipage}
  ~
  \begin{minipage}[c]{0.155\linewidth}
  \includegraphics[width=1\linewidth]{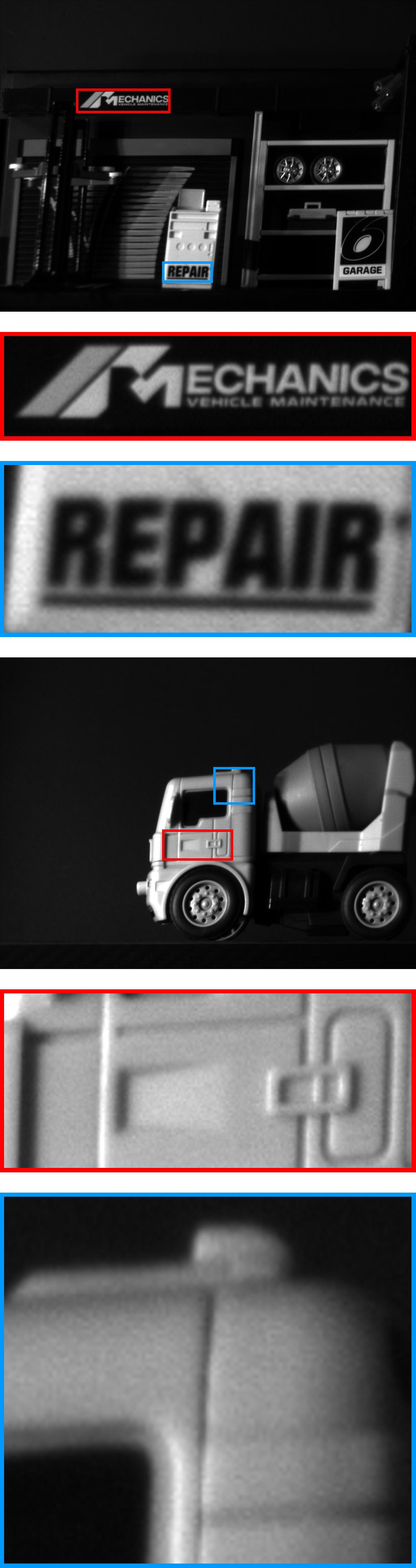}
  \end{minipage}
\end{minipage}
\end{minipage}
\\

   \caption{\textbf{Qualitative evaluations of our method against SOTA SCI image restoration methods on the real dataset capptured by our system in Fig.~\ref{experimental setup}.} Top to bottom shows the results for different scenes. Since the compressed ground truth images in real datasets are unavailable, we capture separate scene images after capture the snapshot compressed image used for reference. For qualitative evaluation purpose, we render images from the learned 3D scene representations by SCINeRF. The results demonstrate that our SCINeRF surpasses existing image restoration methods on real datasets.
   }
   \label{fig5}
   \vspace{-0.7em}
\end{figure*}

\PAR{Results} The experimental results on the synthetic dataset provides empirical evidence on the efficacy of our SCINeRF in estimating and representing high-quality 3D scenes from a single SCI measurement, as shown in Fig. \ref{fig4} and Table \ref{table_2dt3d}. It is noteworthy that, in certain scenes, the SSIM metric of our method does not exceed that of EfficientSCI. This observation can be attributed to the fact that our approach does not directly recover compressed images from the input SCI measurement; instead, it utilizes NeRF to render images based on the underlying scene representation. While NeRF successfully preserves the majority of scene details, the rendering process may introduce a marginal loss of image information. Nevertheless, our method exhibits superior performance in terms of both PSNR and LPIPS metrics. We notice that our SCINeRF performs significantly higher than existing methods in scenes with rich textures and characters, where existing methods fail to retrieve these details as shown in Fig. \ref{fig4}. 

\tabnref{table_3dt3d} presents the quantitative comparisons on novel-view image synthesis. Even in the unfair situation (\ie the other methods exploit camera poses computed by COLMAP \cite{schonberger2016structure} with ground truth images), our method still outperforms existing methods. 

To evaluate the performance of SCINeRF on real datasets, we also conduct qualitative comparisons against SOTA methods. Fig. \ref{fig5} illustrates the experimental results, depicting the outcomes for real datasets. Notably, existing SOTA techniques exhibit limitations when confronted with high-frequency details and characters, resulting in noticeable deficiencies in the output images. In contrast, our SCINeRF surpasses these methods on real datasets by effectively recovering scenes with fine details, thereby achieving superior performance.

\begin{table}
	\centering
	\setlength\tabcolsep{1pt}
	\resizebox{0.85\linewidth}{!}{
		\begin{tabular}{c|ccc}
			\specialrule{0.1em}{1pt}{1pt}
			& \multicolumn{3}{c}{Synthetic Dataset} \\
			\hline
			Mask Overlapping Rate & PSNR$\uparrow$ & SSIM$\uparrow$ & LPIPS$\downarrow$ \\
			\specialrule{0.05em}{1pt}{1pt}
			0.125 & 31.93 & .9562 & .0520 \\
			0.25 & \textbf{33.61} & \textbf{.9599} & \textbf{.0420} \\
			0.5 & 32.92 & .9494 & .0465 \\
			0.75 & 30.08 & .8917 & .1398 \\
			\specialrule{0.1em}{1pt}{1pt}
		\end{tabular}
	}
	\caption{{\bf{Additional studies on the effect of mask overlapping rate.}} The results demonstrate that image quality first increase then decrease when the mask overlapping rate increase. The best mask overlapping rate is 0.25.}
	\label{table_ablation1}
	\vspace{-1.2em}
\end{table}

\subsection{Additional Study}

\PAR{Mask overlapping rate.} We evaluate the effect of different mask overlapping rates when modulating the input frames during SCI image formation process. When capturing the SCI measurement, the camera moves along a trajectory and a series of fixed masks modulate the input multi-view images to form the compressed image. It is possible that there is spatial overlap between masks. We define the mask overlapping rate as:
%
%
\begin{equation}\label{eq_overlapping_rate}
\mathbf{OR}(x,y) = \frac{\sum_{i=1}^{N}\mathbf{M}_i(x,y)}{N},     
\end{equation}
where $\mathbf{OR}$ denotes mask overlapping rate, $\mathbf{M}_i$ is the $i$-th mask and $(x,y)$ indicates pixel coordinate, and $N$ is total number of compressed images. From \eqnref{eq_overlapping_rate}, we know that when mask overlapping rate becomes low, the mask will become sparse in spatial domain, leading to lower sampling rate but higher information loss. However, when overlapping rate becomes high, the measurement values will contain pixels values from more frames. It may lead to pixel value ambiguous and deterioration on the estimated scenes from NeRF. We evaluate the effect of different mask overlapping rates via synthetic datasets. We tested different overlapping rates, ranging from 0.125 to 0.75. As shown in Table \ref{table_ablation1}, when mask overlapping rates increase from 0.125 to 0.25, the image quality becomes higher, indicating the benefit to sample more pixels. However, the image quality decreases rapidly when the overlapping rate continues to increase. It demonstrates that sampling too many pixels would also lead to performance degradation due to the increased ambiguity, which can be better perceived for an extreme case. For example, when the overlapping rate becomes 1, the formation model is the same as that of a blurry image and it is severe ill-posed to recover sharp images from a single blurry image. Empirically, we choose the overlapping rate not to exceed 0.25 for all our experiments.


\PAR{High compression ratio}
We study the performance of our SCINeRF under different compression ratios, i.e., the number of compressed images $N$ within one SCI measurement. High compression ratio is a challenging task for SCI image reconstruction algorithms because the higher compression ratio leads to more information loss, making the image reconstruction and recovery more difficult. In this paper, we test our SCINeRF under the compression ratios of 8, 16, 24 and 32, by running SCINeRF on synthetic \textit{Cozy2room, Tanabata, Factory, Vender} datasets at different compression ratios. Table \ref{table_ablation2} shows the mean metrics of our SCINeRF on four scenes under different compression ratios. It is noteworthy that the image quality of state-of-the-art methods decrease rapidly when compression ratios increase. On the other hand, the image quality deterioration of SCINeRF is small, and our SCINeRF still keeps high image quality even under very-high compression ratios.  

\begin{table}[t]
    \centering
	\setlength\tabcolsep{1pt}
	\resizebox{\linewidth}{!}{
		\begin{tabular}{c|cc|cc|cc}
			\specialrule{0.1em}{1pt}{1pt}
			& \multicolumn{2}{c|}{CR=16} & \multicolumn{2}{c|}{CR=24} & \multicolumn{2}{c}{CR=32}\\
			  & PSNR$\uparrow$ & SSIM$\uparrow$ & PSNR$\uparrow$ & SSIM$\uparrow$ & PSNR$\uparrow$ & SSIM$\uparrow$ \\
			\specialrule{0.05em}{1pt}{1pt}
			GAP-TV & 19.50 & .4975 & 18.63 & .4700 & 18.25 & .4425 \\
			PnP-FFDNet & 24.35 & .7950 & 22.21 & .7325 & 23.55 & .6975 \\
			PnP-FastDVDNet & 26.59 & .8600 & 25.85 & .8475 & 24.56 & .8100 \\
			EfficientSCI & 31.60 & .9093 & 29.10 & .8717 & 28.43 & .8443 \\
			Ours & \textbf{34.88} & \textbf{.9600} & \textbf{34.37} & \textbf{.9532} & \textbf{33.95} & \textbf{.9481} \\
			\specialrule{0.1em}{1pt}{1pt}
		\end{tabular}
	}
	\caption{{\bf{Additional studies on the performance of methods with different compression ratios.}} We test the performance of different methods under the compression ratios (CR) of 16, 24 and 32. The image quality metrics of our SCINeRF decrease slightly when the compression ratio increase.}
	\label{table_ablation2}
	\vspace{-0.8em}
\end{table}


	\section{Conclusion}
\label{sec:con}
In this paper, we present SCINeRF, a novel approach for 3D scene representation learning from a single snapshot compressed image. SCINeRF exploits neural radiance fields as its underlying scene representation due to its impressive representation capability. Physical image formation process of an SCI image is exploited to formulate the training objective for jointly NeRF training and camera poses optimization. Different from previous works, our method considers the underlying 3D scene structure to ensure multi-view consistency among recovered images.   
To validate the effectiveness of SCINeRF, we conduct thorough evaluations against existing state-of-the-art techniques for SCI image recovery with both synthetic and real datasets. Extensive experimental results demonstrate the superior performance of our method in comparison to existing methods, and the necessity to consider the underlying 3D scene structure for SCI decoding. 

\textbf{Acknowledgements.} This work was supported in part by NSFC under Grants 62202389 and 62271414, in part by a grant from the
Westlake University-Muyuan Joint Research Institute, and
in part by the Westlake Education Foundation, Science Fund for Distinguished Young Scholars of Zhejiang Province (LR23F010001), Research Center for Industries of the Future (RCIF) at Westlake University and and the Key Project of Westlake Institute for Optoelectronics (Grant
No. 2023GD007).

	{
		\small
		\bibliographystyle{ieee_fullname}
		\bibliography{bibliography_li}
	}
\end{document}